\documentclass[a4paper,10pt]{article}
\usepackage{amsmath,amssymb,epsfig,graphicx} 
\usepackage{setspace}

\begin{document}


\title{Consistent simulation of non-resonant diphoton production \\
in hadron collisions including associated jet production \\
up to two jets}

\author{Shigeru Odaka\footnote{E-mail: \texttt{shigeru.odaka@kek.jp}.}, 
and Yoshimasa Kurihara\\
High Energy Accelerator Research Organization (KEK)\\
1-1 Oho, Tsukuba, Ibaraki 305-0801, Japan}

\date{}

\maketitle


\begin{abstract}
An event generator for diphoton ($\gamma\gamma$) production 
in hadron collisions that includes associated jet production up to two jets 
has been developed using a subtraction method based on the LLL subtraction.
The parton shower (PS) simulation to restore the subtracted divergent 
components involves both QED and QCD radiation, 
and QED radiation at very small $Q^{2}$ are simulated by referring to 
a fragmentation function (FF).
The PS/FF simulation has the ability to enforce the radiation of 
a given number of energetic photons.
The generated events can be fed to PYTHIA to obtain particle (hadron)-level 
event information, 
which enables us to perform realistic simulations of photon isolation and 
hadron-jet reconstruction.
The simulated events, in which the loop-mediated 
$gg \rightarrow \gamma\gamma$ process is involved, 
reasonably reproduce the diphoton kinematics measured at the LHC. 
Using the developed simulation, 
we found that the 2-jet processes significantly contribute to
diphoton production.
A large 2-jet contribution can be considered as a common feature 
in electroweak-boson production in hadron collisions 
although the reason is yet to be understood. 
Discussion concerning the treatment of the underlying events in photon 
isolation is necessary for future higher precision measurements.
\end{abstract}

\section{Introduction}\label{sec:intro}

The last missing piece of the Standard Model, {\it i.e.}, the Higgs boson, 
was discovered by the ATLAS and CMS experiments at the CERN LHC 
proton-proton ($pp$) collider~\cite{ATLAS:2012gk,CMS:2012gu}.
Diphoton ($\gamma\gamma$) production played a crucial role in this discovery.
The experiments have now entered a new phase that aims to investigate the 
detailed properties of the discovered Higgs boson.
Diphoton production is also an important process 
in this phase~\cite{Khachatryan:2014ira,Aad:2014lwa}.

The Higgs boson was discovered as a resonance in the invariant 
mass distributions of its decay products.
In the diphoton mode, 
the resonance is observed on a large non-resonant diphoton background 
from other processes.
Hence, a precise understanding of this background is indispensable 
for Higgs-boson studies. 
However, our present knowledge is not good enough to predict the background 
properties with sufficient precision.
The background contribution is evaluated using data-driven methods 
in experimental measurements without relying on theoretical predictions 
or simulations.
Because such evaluations are always based on certain assumptions, 
better theoretical understanding is desired for reliable estimations.

Although photons are produced via well-known quantum electrodynamic (QED) 
interactions, it is not straightforward to evaluate the properties 
of diphoton production in hadron collisions.
Experiments at Fermilab Tevatron, a proton-antiproton ($p\bar{p}$) collider 
at a center-of-mass energy ($\sqrt{s}$) of 1.96 TeV, 
observed that next-to-leading order (NLO) 
predictions do not precisely reproduce their measurement 
results~\cite{Abazov:2010ah,Aaltonen:2011via,Aaltonen:2011vk}.
Most remarkably, the CDF experiment found that the NLO predictions 
significantly underestimate the production of acoplanar 
two photons~\cite{Aaltonen:2011via,Aaltonen:2011vk}.
Similar discrepancies have also been observed in experiments 
at the LHC~\cite{Aad:2011mh,Chatrchyan:2011qt}.

Recently, it was demonstrated~\cite{Cieri:2012uk} that the deficit 
of the NLO predictions can be recovered by improving the approximation 
to the next-to-next-to-leading order (NNLO)~\cite{Catani:2011qz}.
A similar improvement was previously found in a simulation with a 
leading-order (LO) event generator, SHERPA~\cite{Gleisberg:2008ta}, 
which includes associated jet production up to two jets~\cite{Hoeche:2009xc}.
Here, a {\it jet} represents a parton (light quark or gluon) in the final state.
These facts imply that the improvements were accomplished predominantly 
by newly included two-jet production processes, even in the NNLO prediction.
More recently, the ATLAS and CMS experiments published their results 
concerning several kinematical distributions of diphoton production 
with significantly improved statistics~\cite{Aad:2012tba,Chatrchyan:2014fsa}.
In these reports, they confirmed the improvements by the NNLO prediction 
and SHERPA simulation.
However, none of the theoretical predictions nor simulations that they examined 
could reproduce all measurement results within the measurement errors, 
which are 10\% to 20\% in most measurement points.
Similar improvements and discrepancies were also observed by the CDF 
experiment~\cite{Aaltonen:2012jd}.

\begin{figure}[tp]
\begin{center}
\includegraphics[scale=0.8]{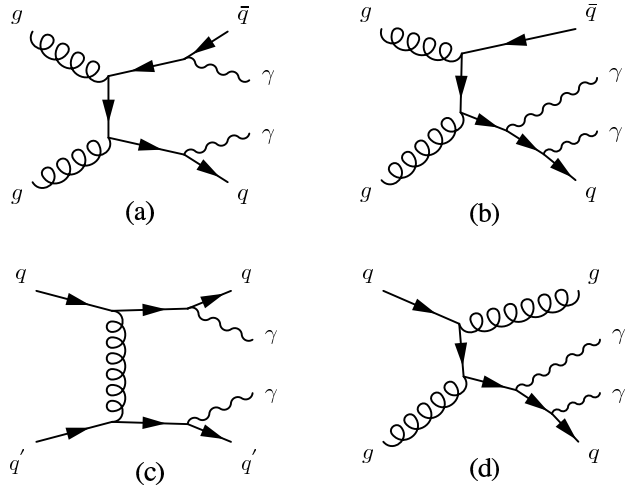}
\caption{\label{fig:diagrams}
Typical Feynman diagrams for non-resonant diphoton production 
in hadron collisions associated with the production of two additional jets.
}
\end{center}
\end{figure}

We reported previously~\cite{Odaka:2012ry} 
that we have successfully developed a consistent Monte Carlo (MC) event 
generator for non-resonant diphoton production processes, 
including those associated with one additional jet.
Using this event generator, 
we confirmed that the contributions of the processes induced 
by constituent quark-gluon ($qg$) collisions overwhelm 
the lowest-order quark-antiquark ($q\bar{q}$) collision processes 
in proton-proton collision experiments, 
owing to the high gluon density inside protons.
This observation further implies that gluon-gluon ($gg$) collisions may 
significantly contribute to diphoton production.
The $gg$ collisions necessarily produce at least two jets in the final state 
in association with diphoton production at the tree level, 
as shown in Figs.~\ref{fig:diagrams}(a) and \ref{fig:diagrams}(b).
In addition, quark-quark scattering processes such as those illustrated in 
Fig.~\ref{fig:diagrams}(c) may also have a large contribution because any 
combination of quarks can contribute.
The inclusion of these qualitatively new processes may be the reason for 
the significant improvements accomplished by the NNLO calculation and 
SHERPA simulation.

The question is now the reason for the remaining discrepancies 
between the measurements and predictions.
In the NNLO predictions, 
the dominant reason may be the contributions from multiple quantum chromodynamic 
(QCD) radiations to be resummed.
In contrast, soft radiation effects are included by parton shower (PS) 
simulations in SHERPA.
Hence, along with missing higher-orders and loop corrections, 
implementation of the radiation-merging method and/or modeling 
are also suspected to be the reason for the inaccuracy of the SHERPA simulation.
Therefore, it must be worth cross-checking the performance of the LO simulations 
with a different merging method and independent implementation.

In this article, we improve 
the MC event generator that we developed previously~\cite{Odaka:2012ry} 
to consistently include diphoton + 2-jet production processes 
by extending our matching method for jet and photon radiations based on the LLL 
subtraction~\cite{Kurihara:2002ne,Odaka:2007gu,Odaka:2009qf,Odaka:2011hc,Odaka:2012da}.
The main difference with the SHERPA simulation is the coverage of PS simulations 
and treatment of non-divergent components.
In the SHERPA simulation, 
which is based on the idea of the CKKW method~\cite{Catani:2001cc}, 
radiation simulations based on matrix-element (ME) calculations are applied 
down to a matching scale ($Q^{2}_{\rm match}$) 
that is significantly smaller than the typical energy scale ($\mu^{2}$) 
of the considered hard interaction for which the MEs are evaluated.
In order to interpolate between the two energy scales, 
the Sudakov suppression deduced from multiple QCD radiation 
is applied to the simulated events by reinterpreting the radiations 
in the context of a PS.
The application of PS simulations is limited in small $Q^{2}$ regions, 
$Q^{2} < Q^{2}_{\rm match}$.
Here, the definition of $Q^{2}$ is model-dependent, 
but at least at the $Q^{2} \rightarrow 0$ limit 
it is equal to $-t$ for spacelike initial-state radiation (ISR) and 
the squared mass of the radiator for timelike final-state radiation (FSR).

In contrast, in our simulation, 
PS simulations that resum divergent logarithmic components to all orders 
are applied up to $\mu^{2}$.
ME-based simulations cover radiation contributions in larger $Q^{2}$ regions 
and those from finite non-logarithmic contributions at $Q^{2} < \mu^{2}$.
Thus, no interpolation is required and the non-logarithmic contributions 
in MEs are preserved in all $Q^{2}$ regions.
As a drawback, 
the boundaries between the ME-based and PS simulations 
for logarithmic components appear in visible regions. 
Therefore, it is crucial to adopt an appropriate kinematics model 
for PS branches.
We demonstrated that the kinematics model that follows the simplest 
massless approximation ($p_{T}$-prefixed kinematics) provides good matching 
between the two simulations~\cite{Odaka:2012ry,Odaka:2007gu,Odaka:2009qf}.

An essentially identical matching method is also applied to QED photon 
radiation from final-state quarks.
In order to accomplish the photon-radiation matching, 
the PS simulation needs to have the ability to radiate photons.
The final-state PS that we previously developed~\cite{Odaka:2012ry} 
simultaneously supports QCD and QED radiation.
In addition, photon radiations with very small $Q^{2}$ that PS simulations 
do not cover are simulated by referring to 
a fragmentation function (FF)~\cite{Bourhis:1997yu}.
In principle, we can produce any number of photons in this PS/FF simulation.
However, most of the produced photons are too soft to be detected. 
The simulation has to be repeated until a hard photon that satisfies 
the detection requirements is produced, in order to simulate 
the photon production in the ordinary way of PS simulations. 
Because the probability of producing a detectable hard photon is very small, 
such a simulation is very inefficient and is practically unrealistic 
if we require two hard photons.

In order to solve this problem, 
the previously developed PS/FF simulation has the ability to enforce 
the radiation of one energetic photon~\cite{Odaka:2012ry}.
This simulation is sufficient for diphoton + 1-jet production processes 
because at least one of the two photons is necessarily well separated 
from the other final-state particles.
However, in diphoton + 2-jet production processes such as those illustrated 
in Fig.~\ref{fig:diagrams}, 
two photons may be collinearly produced with respect to outgoing quarks.
In the present study, 
we extend the simulation so that we can enforce multiple photon radiation.
The developed PS/FF simulation is capable of enforcing the radiation of 
two photons from one outgoing quark (Figs.~\ref{fig:diagrams}(b) and (d)), 
as well as those from the combination of two single radiations 
(Figs.~\ref{fig:diagrams}(a) and (c)).
In principle, we can enforce the radiation of any number of photons 
in all possible combinations. 

In addition to the tree-level processes, 
we also include the loop-mediated $gg \rightarrow \gamma\gamma$ process 
in the present study.
The QCD PS simulations are fully applied to all generated events, 
and the simulated parton-level events are passed to PYTHIA~\cite{Sjostrand:2006za}
in order to obtain the event information at the particle (hadron) level.
Kinematical distributions are derived from the obtained particle-level events.
Photon-isolation cuts and hadron-jet reconstructions can, therefore, be simulated 
realistically.
After checking the internal consistency of the simulation, 
the simulation results are compared with recent measurement results at the LHC 
in order to test the capability of the simulation.

The remainder of this article is organized as follows. 
Section~\ref{sec:strategy} describes the strategy used in our simulation.
Together with the overall strategy, 
the enforced photon radiation in the PS simulation is described.
The internal consistency of the simulation is tested in Section~\ref{sec:matching}. 
After describing the special treatment of the loop-mediated 
$gg \rightarrow \gamma\gamma$ process in Section~\ref{sec:gg2gamgam},
predictions from the simulation are compared with recent measurement results 
in Section~\ref{sec:comp}.
Remarkable features of the diphoton production processes are discussed 
in Section~\ref{sec:discuss} using simulation results.
Finally, discussions are concluded in Section~\ref{sec:concl}.

\section{Simulation strategy}\label{sec:strategy}

\subsection{Matching method}\label{subsec:method}

To simulate events in hadron collisions while allowing associated jet production, 
it is necessary to avoid double counting the jet production in ME calculations 
and PS simulations.
In our matching method, 
the matching is accomplished by numerically subtracting divergent leading-logarithmic 
(LL) terms from the squared MEs of radiative events~\cite{Kurihara:2002ne}.
Because only the LL components are subtracted, 
the finite non-logarithmic components in MEs are preserved in the entire phase space.
The PS simulations applied to corresponding non-radiative events 
restore the subtracted terms without double counting or gaps.
The divergent LL contributions are regularized in PS simulations 
by resumming them to all orders.
Thus, the cross sections to be considered in the simulation are all finite.
Since PS simulations are limited by a certain energy scale ($\mu_{\rm PS}^{2}$), 
the subtraction must also be limited by the same energy scale.
Hence, we call this method the limited leading-log (LLL) subtraction~\cite{Odaka:2007gu}.

We take the above $\mu_{\rm PS}^{2}$ to be equal to the typical energy scale 
($\mu^{2}$) of the hard interaction for which the MEs are evaluated.
Thus, we do not need to account for the Sudakov suppression due to 
the multiple QCD radiation in the ME evaluation.
Because we take $\mu_{\rm PS}^{2}$ be large, 
the boundaries between the ME-based simulation and PS simulation 
for the LL contributions emerge in visible regions. 
Hence, special care is required in both simulations 
in order to achieve good matching between them.
On the ME side, in the subtraction procedure, 
a radiative event is separated into a hypothetical PS branch and a non-radiative event 
in which the assumed radiation is removed by exactly reversing the PS procedure.
The energy scale that limits the subtraction is determined by referring to 
the PS scale that is defined for the reconstructed non-radiative event. 
The subtraction is applied only when the $Q^{2}$ of the separated PS branch 
is smaller than the obtained PS scale.
Here, the subtraction functions are tailored so that they should be exactly identical 
to the leading term of the PS simulation~\cite{Odaka:2011hc,Odaka:2012ry}.

On the PS side, 
the modeling of PS branch kinematics is important for the matching.
We adopt the $p_{T}$-prefix kinematics in which the $p_{T}$ value of each branch 
is fixed to the value that is determined from the simplest massless approximation; 
{\it i.e.}, $p_{T}^{2} = (1-z)Q^{2}$ for ISR~\cite{Odaka:2009qf} and 
$p_{T}^{2} = z(1-z)Q^{2}$ for FSR~\cite{Odaka:2011hc}.
In order to realize this kinematics, 
the identity of $Q^{2} = -t$ is allowed to be violated in ISR~\cite{Odaka:2007gu}, 
and the energy conservation is violated in FSR~\cite{Odaka:2012da}.
The energy conservation is restored by adjusting the initial-state momenta 
after completing the PS simulations.
Accordingly, some corrections are applied to compensate for the adjustment 
in the initial state~\cite{Odaka:2012ry}.

We execute the LLL subtraction and PS simulations in a frame 
where incoming partons are aligned back to back.
In such a setup, 
we do not need to separately consider the soft-gluon (SG) divergence 
when the studied process includes only one jet in the final state.
However, when two or more jets are included, 
it is necessary to consider the SG divergences together with the collinear 
divergences that are subtracted by the LLL subtraction 
to render the cross sections finite.
In our simulation, the SG divergences are subtracted simultaneously with 
collinear divergences using the combined subtraction method~\cite{Odaka:2014ura}.
In order to compensate for this alteration in the subtraction, 
a correction is applied to the events of corresponding non-radiative processes 
by referring to the hardest branch in the PS simulation (SG correction)~\cite{Odaka:2014ura}.

A similar matching method is also applied to QED photon radiation 
in photon production processes~\cite{Odaka:2012ry,Odaka:2015uqa}.
The cross section diverges if a photon and a quark are produced collinearly.
Although such a photon is likely to be confined in a hadron jet induced by the quark, 
the photon may be detected as an isolated photon when the quark momentum is small.
The LLL subtraction is applied to regularize such a final-state divergence,  
whereas initial-state divergences and soft-photon divergences are not taken 
into consideration 
because we always require the detection of energetic photon(s) at large angles 
when we simulate photon production processes.

In order to restore the subtracted QED LL components, 
our final-state PS simulation has the ability to radiate photons 
in the same manner as QCD parton radiation~\cite{Odaka:2012ry}.
PS simulations are necessarily cutoff at a small $Q^{2}$ to avoid divergences.
We terminate the PS simulation at $Q_{0}^{2} = (5 {\rm ~GeV})^{2}$.
Although QCD effects at smaller $Q^{2}$ can be simulated down to the particle level 
with the help of general-purpose event generators such as PYTHIA, 
similar QED simulations are not available in these generators.
In our PS simulation, 
photon radiations at $Q^{2} < Q_{0}^{2}$ are simulated by referring to 
a fragmentation function (FF)~\cite{Bourhis:1997yu}.
In addition, in order to improve the simulation efficiency, 
the PS/FF simulation is capable of enforcing the photon radiation.
Although the original version~\cite{Odaka:2012ry} could radiate only one photon 
in each event, the simulation has been extended to multiple photon radiations, 
as described later.

\subsection{Matched diphoton + 2-jet production}\label{subsec:matched}

Programs for calculating the MEs of diphoton ($\gamma\gamma$) + 2-jet production 
processes were produced using the GRACE system~\cite{Ishikawa:1993qr,Yuasa:1999rg} 
and installed in the GR@PPA event generator~\cite{Odaka:2011hc}.
The $\gamma\gamma$ + 2-jet production ($aa2j$) MEs have various divergences 
to be subtracted.
Together with the initial-state divergences, 
we have to subtract the final-state QCD collinear divergences 
that emerge when the two jets are produced collinearly.
Furthermore, in addition to the collinear divergences, 
we also have to consider the SG divergences when a gluon is included in the final state.
This divergence is subtracted simultaneously with the collinear divergences 
using the combined subtraction method~\cite{Odaka:2014ura}.

The subtracted QCD divergences are restored by combining $\gamma\gamma$ + 1-jet 
production ($aa1j$) events, to which QCD PS simulations are applied.
Here, the SG correction is applied in order to compensate for the alteration 
in the subtraction from $aa2j$ due to the SG divergence.
Because $aa1j$ also has the initial-state QCD divergence, 
the LLL subtraction is applied and $\gamma\gamma$ + 0-jet ($aa0j$) events are combined 
to restore the subtracted components, as in the previous study~\cite{Odaka:2012ry}.

The final-state QED divergences are also subtracted to make the $aa2j$ cross sections finite. 
The subtracted QED components are restored by combining $\gamma$ + 2-jet ($a2j$) events.
The $a2j$ events again have both QCD and QED divergences.
The QCD divergences are subtracted by recursively applying the combined subtraction 
and are restored by combining $\gamma$ + 1-jet production ($a1j$) events 
to which the SG correction is applied. 
The QED divergences are subtracted using the LLL subtraction and restored 
by combining QCD 2-jet ($qcd2j$) events, 
as in the simulation of direct-photon production~\cite{Odaka:2015uqa}.
The radiation of one energetic photon is enforced in the PS simulation 
in the generation of $a2j$ and $a1j$ events, 
and the radiation of two energetic photons is enforced in the $qcd2j$ event generation.
The QED divergences are also subtracted from the $aa1j$ events and  
restored by the forced photon radiation from $a1j$ events.

When one of the four particles in an $aa2j$ event is judged to be a soft radiation 
($Q^{2} < \mu_{PS}^{2}$) for which subtraction should be considered, 
the reconstructed non-radiative three-body event may again contain a soft radiation.
The subtraction is not applied to such {\it doubly-soft} events.
Instead, they are rejected.
Namely, the non-divergent components in {\it doubly-soft} contributions are 
ignored~\cite{Odaka:2014ura}.
In addition to the LL components,  
we further need to subtract next-to-leading logarithmic (NLL) components 
in order to evaluate these non-divergent components.
The evaluation of NLL components is beyond the scope of the present study.
Nevertheless, dominant LL components in the {\it doubly-soft} configuration are 
restored by the PS simulations applied to $aa0j$, $a1j$, and $qcd2j$ events. 
Moreover, some higher-order components are restored by PS simulations applied to 
$aa1j$ and $a2j$ events that contain a non-divergent soft radiation.

In total, the events of six processes, $aa2j$, $aa1j$, $aa0j$, $a2j$, $a1j$, 
and $qcd2j$, are generated according to their tree-level MEs.
The combined subtraction is applied to $aa2j$ and $a2j$, 
and the LLL subtraction is applied to $aa1j$.
One-photon radiation is enforced in the generation of $a2j$ and $a1j$, 
and two-photon radiation is enforced in $qcd2j$.
In addition, the SG correction is applied to $aa1j$ and $a1j$.
All these events are combined to compose a consistent simulation of diphoton production 
allowing associated jet production up to two jets.
We take the gluon and light quarks up to the bottom quark 
as the initial-state partons and final-state jets in the event generation.

In order to carry out the event generation, 
we need to explicitly define the energy scale $\mu$ of hard interactions.
Although any definition is in principle acceptable in our matching method, 
we adopt the following definition as in the previous study~\cite{Odaka:2015uqa}; 
\begin{equation}\label{eq:scale}
  \mu = {\rm max}\{ Q_{T,i} \} .
\end{equation}
Here, quantity $Q_{T,i}$ is defined for each final-state particle 
in terms of its mass and $p_{T}$, such that
\begin{equation}\label{eq:qt}
  Q_{T,i}^{2} = m_{i}^{2} + p_{T,i}^{2} , 
\end{equation}
and the largest value is taken as the energy scale of the event.
This definition reduces to the $p_{T}$ value of the particle that has the largest $p_{T}$, 
as all particles are nearly massless in the present study.

In the event generation, we are required to specify several energy scales, such as 
the renormalization scale ($\mu_{R}$) to determine the coupling parameters in 
ME calculations and the factorization scale ($\mu_{F}$) to resum the initial-state 
QCD radiations in parton-distribution functions (PDFs).
Furthermore, boundary $\mu_{\rm PS}$ for the LL components described above 
may be independently defined for ISR and FSR ($\mu_{\rm ISR}$ and $\mu_{\rm FSR}$).
In the present study, all these energy scales are taken to be equal to $\mu$, 
which is defined in Eq.~(\ref{eq:scale}).

The parton-level events to which the PS simulations are fully applied in GR@PPA 
are passed to PYTHIA 6.425~\cite{Sjostrand:2006za} 
to simulate the hadronization and particle decays. 
Small-$Q^{2}$ QCD effects at $Q^{2} < (5 {\rm ~GeV})^{2}$ that are not covered 
by the GR@PPA PS are also simulated in PYTHIA.
The PYTHIA simulation is applied with its default setting, 
except for the settings of {\tt PARP(67) = 1.0} and {\tt PARP(71) = 1.0},
as in all our previous studies~\cite{Odaka:2009qf,Odaka:2011hc}.
The event selection is applied to the obtained particle-level events to derive 
kinematical distributions, although the diphoton kinematics are not significantly 
affected by this PYTHIA simulation.

The kinematical conditions in the hard-interaction generation and the cuts to select 
the events to be passed to PYTHIA are sufficiently relaxed 
in order not to bias the final particle-level selection.
Among them, the generation conditions are markedly relaxed 
because the PS simulations significantly alter the event kinematics.
In order to ensure efficient event generation under such conditions, 
we adopt the LabCut framework~\cite{Odaka:2012ry} for the generation of all processes.
In this framework, 
the PS simulations and event selection are applied before 
passing the differential cross-section values 
to the MC integration and event generation utility 
BASES/SPRING~\cite{Kawabata:1985yt,Kawabata:1995th} used in GR@PPA.
Hence, the distribution of random numbers for the hard-interaction generation is 
automatically optimized accounting for the event selection conditions 
after the PS application.
In addition, event weights from the SG correction and forced photon radiation 
can be involved in the differential cross-section values using this framework.

Usually, kinematical cuts are merely applied to the photons in diphoton studies. 
Accordingly, we impose practically no constraint on the jets in the 
hard-interaction events.
Even when some requirements are imposed on hadron jets in a study, 
it is dangerous to apply corresponding cuts to the jets (partons) in hard-interaction events 
because additional hadron jets may be produced by PS simulations.
Exceptions are small cutoffs in the $p_{T}$ values and $\Delta R$ separation to 
the other particles, {\it i.e.}, $p_{T} > 1$ GeV/$c$ and $\Delta R > 0.01$, 
where $p_{T}$ is measured with respect to the beam axis and $\Delta R$ is 
defined by the separations in the azimuthal angle ($\phi$) and pseudorapidity ($\eta$) 
as $\Delta R^{2} = \Delta\phi^{2} + \Delta\eta^{2}$.
These cutoffs are applied to ensure numerical stability of the subtraction.
This very loose setting is allowed because all divergences are precisely subtracted 
from the MEs of radiative processes.

\subsection{Forced multiple photon radiation}\label{subsec:photorad}

In the present study, the forced photon radiation in the final-state PS simulation 
developed in a previous study~\cite{Odaka:2012ry} 
is extended to multiple photon radiation.
The procedure is essentially the same as that for single radiation.
The difference is mainly in the determination of the $Q^{2}$ values of photon radiation.
In the PS simulation, we first determine which quark should radiate 
what number of photons.
This is randomly determined by assuming an equal probability for all combinations.
Here, we do not consider any photon radiation from gluons.
The number of possible combinations of this assignment can be given 
by the homogeneous product.
Provided there are $n$ quarks in the final state, 
the number of ways to assign $m$ photons to them is given as
\begin{equation}\label{eq:num_comb}
	N_{\rm comb} = H(n,m) = {(n+m-1)! \over (n-1)!m!} .
\end{equation}

Once the number of photons to radiate is determined for each quark, 
we can determine the $Q^{2}$ values of the photon radiations in decreasing order 
using the QED Sudakov form factor $S_{{\rm QED}}(Q_{1}^{2},Q_{2}^{2})$~\cite{Odaka:2012ry}.  
The $Q^{2}$ values are determined by solving
\begin{equation}\label{eq:q2solv}
	S_{{\rm QED}}(Q^{2},Q_{\rm pre}^{2}) = \xi ,
\end{equation}
where $Q_{\rm pre}^{2}$ is the $Q^{2}$ of the previous radiation and 
is equal to $\mu_{\rm PS}^{2}$ for the first radiation.
In ordinary PS simulations, $\xi$ is a uniform random number between 0 and 1, 
whereas in the forced radiation, $\xi$ is restricted in the range 
\begin{equation}\label{eq:etarange}
	S_{{\rm QED}}(Q_{0}^{2},Q_{\rm pre}^{2}) < \xi < 1 ,
\end{equation}
in order to ensure that we always obtain a solution in the range 
$Q_{0}^{2} < Q^{2} < Q_{\rm pre}^{2}$. 
Here, $Q_{0}^{2}$ is the lower cutoff of the PS simulation.
This constraint leads to an event weight of 
\begin{equation}\label{eq:q2weight}
	w = 1 - S_{{\rm QED}}(Q_{0}^{2},Q_{\rm pre}^{2}) .
\end{equation}
Because the Sudakov form factor represents the no-radiation probability, 
the event weight in Eq.~(\ref{eq:q2weight}) corresponds to the probability of 
obtaining a radiation in the relevant range.

The photon radiations at $Q^{2} < Q_{0}^{2}$ are covered by the FF-based simulation.
The FF-based radiation is considered only for the last radiation from each quark.
Therefore, the event weight in Eq.~(\ref{eq:q2weight}) is modified to 
\begin{equation}\label{eq:q2weight1}
	w_{\rm last} = \left\{1 - S_{{\rm QED}}(Q_{0}^{2},Q_{\rm pre}^{2})\right\} 
	+ \left\{1 - e^{-P_{\rm FF}(Q_{0}^{2})}\right\} 
\end{equation}
for the last radiation, where $P_{\rm FF}(Q_{0}^{2})$ is the integration of 
the radiation probability density given by the FF.
We determine whether the photon is radiated by the PS or FF-based simulation 
according to the ratio of the radiation probabilities described by the first 
and second terms in Eq.~(\ref{eq:q2weight1}).
If PS is selected, the $Q^{2}$ value is determined by solving the equation 
in Eq.~(\ref{eq:q2solv}) with the constraint in Eq.~(\ref{eq:etarange}).

The base event-weight is determined by the product of the number of possible combinations 
given in Eq.~(\ref{eq:num_comb}) and the weights in Eq.~(\ref{eq:q2weight}) or 
Eq.~(\ref{eq:q2weight1}) evaluated for all photon radiations.
The subsequent procedure is the same as that for the single 
forced-radiation~\cite{Odaka:2012ry}.
Photon radiations are inserted during the QCD evolution when the $Q^{2}$ becomes 
smaller than the predetermined photon-radiation $Q^{2}$.
A multiple insertion happens if multiple photon-radiations are assigned to a quark.
An FF radiation is added after completing the QCD evolution 
if it is selected for the last radiation.
The splitting parameter $z$, which determines the momenta of the branch products, 
is determined during this procedure.
The restrictions on the allowed $z$ range to ensure sufficient energy 
for the radiated photons determine the additional event weights to be multiplied.
The event weight is set to zero if the quark does not have sufficient energy.

\begin{figure}[tp]
\begin{center}
\includegraphics[scale=0.6]{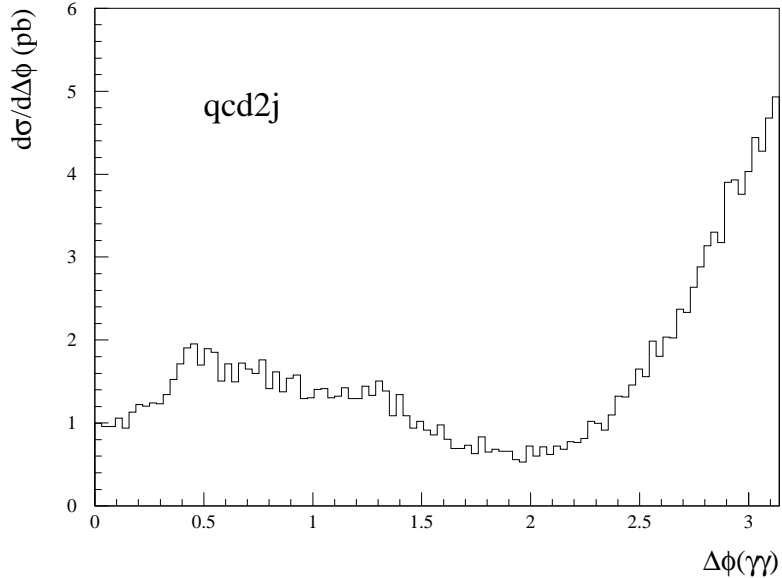}
\caption{\label{fig:dphi-qcd2j}
$\Delta\phi$ distribution of the two photons forced to be radiated in the PS simulation 
applied to $qcd2j$ events.
The events were simulated under the condition for the ATLAS measurement.}
\end{center}
\end{figure}

The differential cross section values to be passed to BASES/SPRING are multiplied 
with the thus-determined event weight using the LabCut framework of GR@PPA, 
for appropriate cross-section integration and the generation of 
unweighted events.
The number of forced photon-radiations ($N_{\gamma}$) and minimum energy 
($E_{\rm min}$) for the photon radiation can be specified by users 
in the PS simulation developed in this study.
Although the $E_{\rm min}$ cut is applied in the PS simulation that is executed 
in the center-of-mass frame of the hard interaction, 
it does not bias the generated events if its value is set to the minimum $p_{T}$ 
requirement for the photons.

Diphoton events can be generated by applying this PS simulation to $qcd2j$ events 
by setting $N_{\gamma} = 2$.
Figure~\ref{fig:dphi-qcd2j} shows the distribution of the azimuthal angle separation 
($\Delta\phi$) between the radiated two photons.
The events were simulated under the condition imposed in the comparison 
with the ATLAS measurement, which is described later.
We can clearly observe two different components in this simulation; 
the two single-radiation contributions such as those corresponding to the collinear 
components of the processes in Figs.~\ref{fig:diagrams}(a) and (c) concentrate 
in a coplanar region, $\Delta\phi \gtrsim 2$, 
while the double-radiation contributions such as those corresponding to the processes 
in Figs.~\ref{fig:diagrams}(b) and (d) emerge in an acoplanar region, 
$\Delta\phi \lesssim 2$.
Events at $\Delta\phi < 0.4$ are suppressed by the required separation 
between the two photons.

\section{Matching test}\label{sec:matching}

The internal consistency of the radiation matching can be tested 
by investigating the stability of kinematical distributions of the simulated events 
against the variation of the energy scale.
Here, we take the $\mu$ value given by the definition in Eq.~(\ref{eq:scale}) 
as the standard value $\mu_{0}$ and compare the distributions for 
the settings of $\mu = 0.5 \mu_{0}$, $\mu_{0}$, and $1.5 \mu_{0}$.
Although usually the larger scale is set to $\mu = 2 \mu_{0}$ in scale-dependence studies, 
the choice of very large $\mu$ value is troublesome in our simulation 
because large PS activities may boost small-$p_{T}$ photons into detectable regions.
Such small-$p_{T}$ photons cannot be properly simulated
because the initial-state QED divergences and soft-photon divergences 
are not subtracted.
This problem is severe in large multiplicity processes 
such as $\gamma\gamma$ + 2-jet production.
In any case, $\mu = 1.5 \mu_{0}$ is large enough to test the stability around 
$\mu = \mu_{0}$.
The kinematical distributions are examined in terms of the normalized differential cross section 
$(1/\sigma)d\sigma/dx$, where $x$ represents the tested quantity,
because the overall normalization is not relevant to the matching properties.

\begin{figure}[tp]
\begin{center}
\includegraphics[scale=0.6]{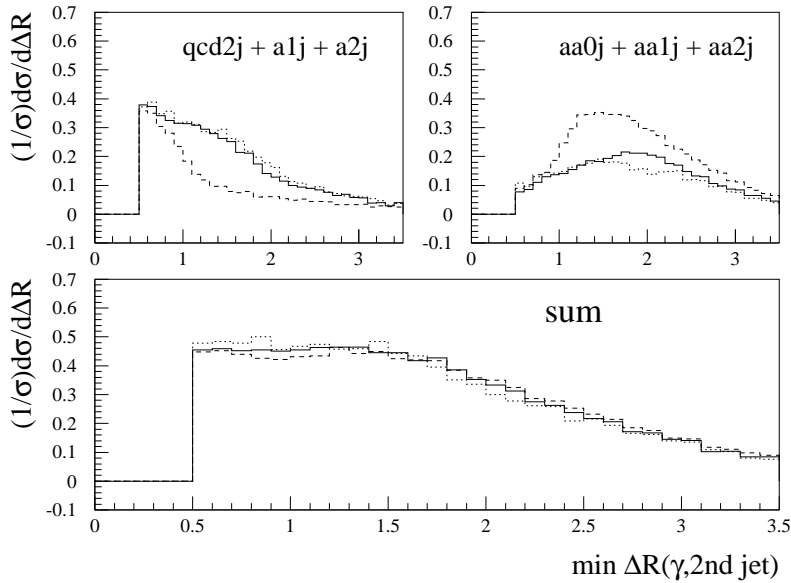}
\caption{\label{fig:drgamj2}
Distribution of the $\Delta R$ separation between the photon and second jet 
in the events for the matching test.
The $\Delta R$ value was evaluated for both photons and the smaller one was histogramed.
The sum of the 0-$\gamma$ and 1-$\gamma$ contributions and the 2-$\gamma$ contribution 
are separately presented in the upper panels, 
and the total sum is presented in the lower panel.
The dashed, solid, and dotted histograms represent the results for $\mu = 0.5\mu_{0}$, 
$\mu_{0}$, and $1.5 \mu_{0}$, respectively.
}
\end{center}
\end{figure}

The event simulation was carried out for $pp$ collisions at $\sqrt{s}$ = 7 TeV 
using the built-in CTEQ6L1 PDF~\cite{Pumplin:2002vw}.
We deactivated the underlying-event (UE) simulation in PYTHIA by setting {\tt MSTP(81) = 0} 
because it does not affect the properties of energetic objects to be examined 
in the following procedure.
From the simulated particle-level events, 
we selected those events in which the two generated photons satisfied the conditions  
\begin{equation}\label{eq:photon1}
  p_{T}({\gamma}) > 100 {\rm ~GeV}/c , {\rm ~~} |\eta({\gamma})| < 2.4 ,
  {\rm ~~and~~} \Delta R(\gamma\gamma) > 0.4 , 
\end{equation}
where $p_{T}$ is measured with respect to the beam direction and 
$\Delta R(\gamma\gamma)$ represents the $\Delta R$ separation between the two photons.
In addition, both photons were required to be isolated from other activities 
with the condition of 
\begin{equation}\label{eq:phiso1}
  E_{T}^{\rm cone}(\Delta R < 0.4) < 7 {\rm ~GeV} 
\end{equation}
in order to simulate a realistic detection condition.
Cone $E_{T}$, $E_{T}^{\rm cone}(\Delta R < 0.4)$, is defined by the sum of 
the transverse energies of all stable particles other than muons and neutrinos 
that are contained inside a cone of $\Delta R = 0.4$ around the photon. 

We examine the properties of $\gamma\gamma$ + 2-jet events.
Here, {\it jet} does not refer to a parton but an observable hadron jet.
The hadron jets were reconstructed using FastJet 3.0.3~\cite{Cacciari:2011ma}, 
with the application of the anti-$k_{T}$ algorithm with $R = 0.4$.
All stable particles within $|\eta| < 5.0$, including neutrinos, 
were used for the reconstruction.
The reconstructed jets that satisfied the conditions 
\begin{equation}\label{eq:jet1}
  p_{T}({\rm jet}) > 30 {\rm ~GeV}/c ,  ~~|\eta({\rm jet})| < 4.4 ,  
  {\rm ~~and~~}\Delta R(\gamma, {\rm jet}) > 0.5 
\end{equation}
were taken as the detected jets, 
where $\Delta R(\gamma,{\rm jet})$ is the separation in $\Delta R$ 
between the photon and jet.
We required the detection of two jets in each event.
The large $p_{T}$ threshold for the photons in Eq.~(\ref{eq:photon1}) 
was adopted so that hadron jets sufficiently softer than the energy scale of the event 
are always allowed to be produced.
Such soft jets are predominantly produced by PS simulations.

Figure~\ref{fig:drgamj2} shows the distribution of the $\Delta R$ separation 
between the photon and second jet, {\it i.e.}, the detected hadron jet having 
the second largest $p_{T}$.
The smaller of the two values evaluated for both photons is histogramed in the figure.
The sum of the results from the 0-$\gamma$ ($qcd2j$) and 1-$\gamma$ ($a2j$ + $a1j$) processes 
is presented in the upper-left panel, 
and that from the 2-$\gamma$ processes ($aa2j$ + $aa1j$ + $aa0j$) is presented 
in the upper-right panel.
The lower panel shows the total sum of the two results, {\it i.e.}, 
the sum of the results from all six processes.
The dashed, solid, and dotted histograms illustrate the results for the settings 
of $\mu = 0.5\mu_{0}$, $\mu_{0}$, and $1.5\mu_{0}$, respectively.

We expected this distribution to be sensitive to the photon radiation from 
the final-state quark, as in the $\gamma$ + 2-jet production~\cite{Odaka:2015uqa}.
Indeed, an increase of large-angle photon radiations in the 0-$\gamma$ + 1-$\gamma$ result 
and a corresponding decrease in the 2-$\gamma$ result are clearly observed 
when we increase the energy scale from $0.5\mu_{0}$ to $\mu_{0}$. 
However, a similar effect is not clear when the scale is increased from $\mu_{0}$ 
to $1.5\mu_{0}$.
This is because $\mu$ defines the boundaries for both QED and QCD radiations, 
and its effect is complicated for large $\mu$ values. 
Substantially, the summed distribution is very stable against the $\mu$ variation, 
implying good matching between the ME-based and PS simulations. 

\begin{figure}[tp]
\begin{center}
\includegraphics[scale=0.6]{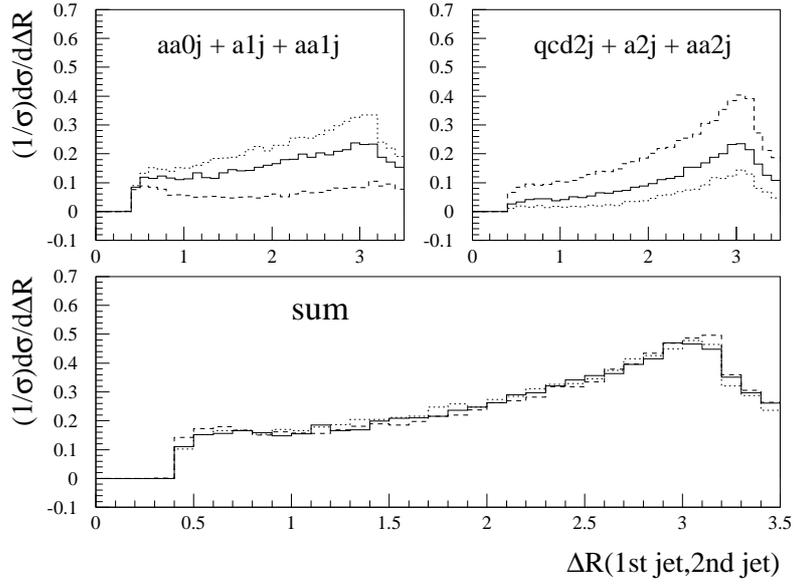}
\caption{\label{fig:drj1j2}
Distribution of the $\Delta R$ separation between the leading jet and second jet 
in the events for the matching test.
The sum of the 0-jet and 1-jet contributions and the 2-jet contribution are 
separately presented in the upper panels, 
and the total sum is presented in the lower panel.
The dashed, solid, and dotted histograms represent the results for $\mu = 0.5\mu_{0}$, 
$\mu_{0}$, and $1.5 \mu_{0}$, respectively.
}
\end{center}
\end{figure}

\begin{figure}[tp]
\begin{center}
\includegraphics[scale=0.6]{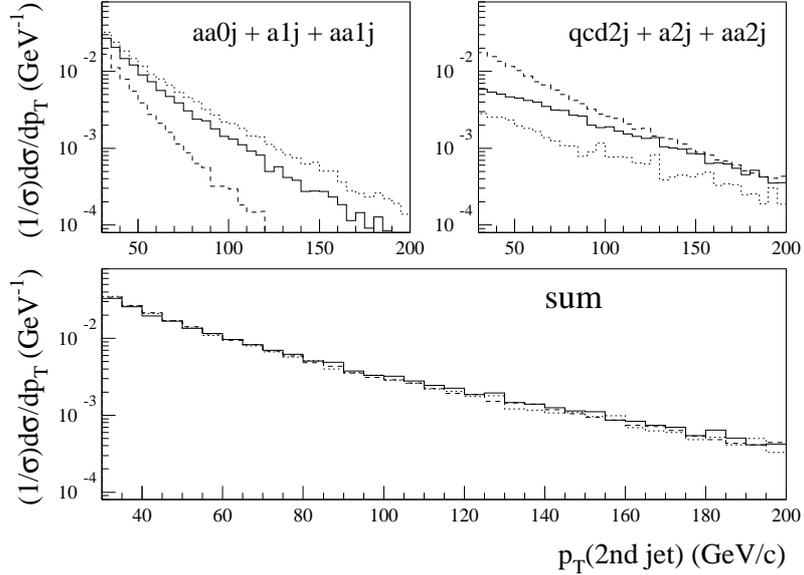}
\caption{\label{fig:ptjet2}
$p_{T}$ distribution of the second jet.
The sum of the 0-jet and 1-jet contributions and the 2-jet contribution are 
separately presented in the upper panels, 
and the total sum is presented in the lower panel.
The dashed, solid, and dotted histograms represent the results for $\mu = 0.5\mu_{0}$, 
$\mu_{0}$, and $1.5 \mu_{0}$, respectively.
}
\end{center}
\end{figure}

The results for the $\Delta R$ separation between the leading (largest-$p_{T}$) jet 
and second jet are presented in Fig.~\ref{fig:drj1j2}, 
and those for the second-jet $p_{T}$ are presented in Fig~\ref{fig:ptjet2}.
In both figures, the sum of the results from the 0-jet ($aa0j$) and 1-jet ($aa1j$ + $a1j$) 
processes is shown in the upper-left panel, 
that from the 2-jet ($aa2j$ + $a2j$ + $qcd2j$) processes in the upper-right panel, 
and the total sum is shown in the lower panel.
The other notations are same as those in Fig.~\ref{fig:drgamj2}.
We expected the distribution in Fig.~\ref{fig:drj1j2} to be sensitive to 
the final-state QCD radiation and the distribution in Fig.~\ref{fig:ptjet2} to be 
sensitive to the initial-state QCD radiation.
Although the observations are less clear than those in $\gamma$ + 2-jet production, 
we can see a general tendency for the contribution from smaller jet-multiplicity 
processes to increase as the energy scale increases, 
and, accordingly, the contribution from the 2-jet processes decreases.
The resultant large energy-scale dependences in the separate results compensate 
for each other to result in stable distributions in the summed results.
These observations again confirm the good matching property of the simulation.

\section{Loop-mediated $gg \rightarrow \gamma\gamma$ process}\label{sec:gg2gamgam}

\begin{figure}[tp]
\begin{center}
\includegraphics[scale=0.7]{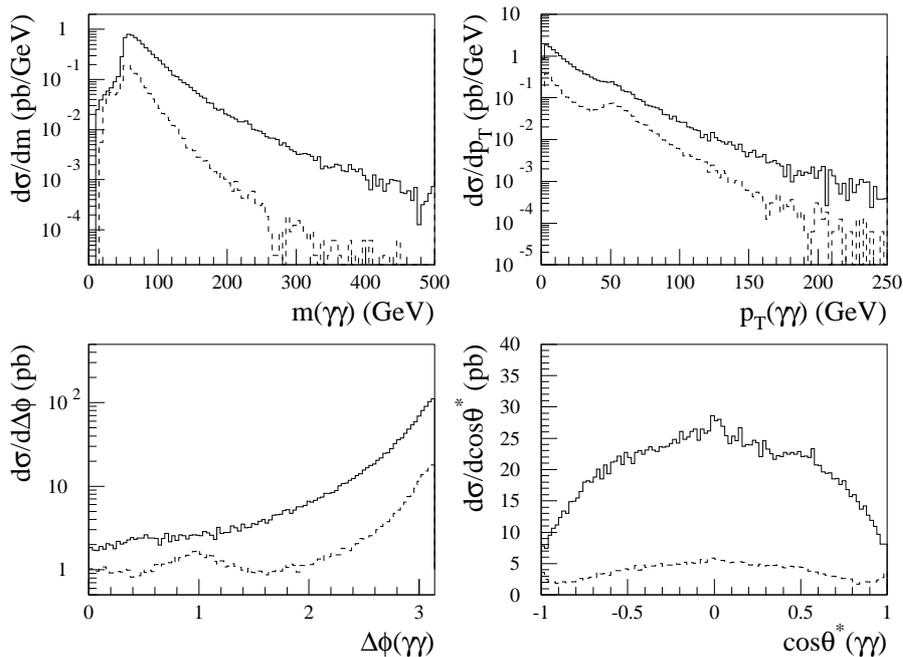}
\caption{\label{fig:gg2aa}
Contribution of the $gg \rightarrow \gamma\gamma$ process to the diphoton kinematical 
distributions.
The simulation was carried out under the condition for the ATLAS measurement.
The $gg \rightarrow \gamma\gamma$ results (dashed histograms) are compared with 
those of the tree-level matched simulation (solid histograms) described 
in Section~\ref{sec:strategy}.
}
\end{center}
\end{figure}

It is known that the contribution from the fermion-loop mediated 
$gg \rightarrow \gamma\gamma$ process to the diphoton production in $pp$ collisions 
is unignorable because of the high gluon density inside protons.
We implement this process in GR@PPA by hand-coding the differential cross section 
formula in a literature~\cite{Berger:1983yi}.
Thus, the lowest-order contribution can be included in the simulation. 
However, some of the kinematical distributions that are compared with the simulation 
in later sections are sensitive to extra jet radiations.
We simulate such radiation effects by adopting a large energy scale in the event generation.
We can choose a very large value because this process is independent of the tree-level 
matched processes described in Section~\ref{sec:strategy}. 
In the present study, we set 
\begin{equation}\label{eq:scale_gg2aa}
  \mu = 2 m_{\gamma\gamma}
\end{equation}
for the generation of $gg \rightarrow \gamma\gamma$ events, 
where $m_{\gamma\gamma}$ is the invariant mass of the produced two photons.
The event generation is carried out in the same manner as the tree-level matched 
processes using the LabCut framework, 
and the generated events are processed by PYTHIA to obtain the particle-level event 
information.

Because $\mu$ defines the upper limit of the PS simulation, 
the QCD radiations from initial-state partons are simulated up to 
$Q^{2} = (2 m_{\gamma\gamma})^{2}$ according to the LL approximation in PS 
with the above setting.
This simulation should not be far away from reality 
because $gg \rightarrow \gamma\gamma$ has a good perturbative property, {\it i.e.}, 
the NLO correction is reasonably small~\cite{Nadolsky:2007ba}.
Indeed, our simulation reproduces the $p_{T}$ distribution of the diphoton system 
evaluated by a resummed NLO calculation~\cite{Nadolsky:2007ba} reasonably well.

The dashed histograms in Fig.~\ref{fig:gg2aa} show the $gg \rightarrow \gamma\gamma$ 
($gg2aa$) contribution to the kinematical distributions of the two photons 
measured by the ATLAS experiment~\cite{Aad:2012tba}. 
The simulation was carried out with the above setting under the condition for 
simulating the measurement that is described later.
The simulation results are compared with those of the tree-level matched 
simulations illustrated with solid histograms.
The $gg2aa$ contribution is approximately 16\% of the total yield. 
Although this contribution is substantial in small $m(\gamma\gamma)$ and 
small $\Delta\phi(\gamma\gamma)$ regions, 
it provides only minor corrections to the distributions in most regions.

\section{Comparison with LHC measurements}\label{sec:comp}

\subsection{ATLAS measurement}\label{subsec:atlas}

The ATLAS experiment published their measurement results regarding the kinematical 
distributions of two photons produced in $pp$ collisions~\cite{Aad:2012tba}.
The measurement is based on 4.9 ${\rm fb}^{-1}$ data at $\sqrt{s}$ = 7 TeV 
provided by the LHC.
We simulated this measurement following the signal definition described in their 
report.

The simulation was carried out using the MRST2007LO* PDF~\cite{Sherstnev:2007nd} 
because this PDF is widely used for LO simulations in experimental studies.
The two generated photons were required to be observed in angular ranges of 
\begin{equation}\label{eq:phang_atlas}
  |\eta({\gamma})| < 1.37 {\rm ~~or~~} 1.52 < |\eta({\gamma})| < 2.37 
\end{equation}
after completing the simulation down to the particle level.
An asymmetric $p_{T}$ requirement of 
\begin{equation}\label{eq:phmom_atlas}
  p_{T}(\gamma_{1}) > 25 {\rm ~GeV}/c {\rm ~~and~~} p_{T}(\gamma_{2}) > 22 {\rm ~GeV}/c
\end{equation}
and an angular separation requirement of
\begin{equation}\label{eq:phsepa_atlas}
  \Delta R(\gamma\gamma) > 0.4
\end{equation}
were imposed on the observed two photons. 
In addition, an isolation condition of 
\begin{equation}\label{eq:phiso_atlas}
  E_{T}^{\rm cone}(\Delta R < 0.4) < 4 {\rm ~GeV} 
\end{equation}
was required for both photons.
The definition of $E_{T}^{\rm cone}(\Delta R < 0.4)$ is same as that 
in Eq.~(\ref{eq:phiso1}).
The UE simulation in PYTHIA was deactivated as in Section~\ref{sec:matching} 
because its effect is subtracted from $E_{T}^{\rm cone}$ in the measurement 
together with the pile-up effects.
Contrary to the simulation in Section~\ref{sec:matching}, 
no constraint was imposed on the hadron jets that may be produced in association 
with the two photons.

\begin{table}
\begin{center}
\caption{
Contribution of each process in the simulation of the ATLAS measurement. 
Events of weight $+1$ or $-1$ are generated according to the folded 
cross section ($\sigma_{\rm abs}$) and 
$R_{\rm neg}$ represents the fraction of negative-weight events.
The cross section ($\sigma$) is obtained from the difference between the 
numbers of positive- and negative-weight events.
}
\label{table:atlas}
\begin{tabular}{lcccc}
\hline
process & $\sigma_{\rm abs}$ (pb) & $R_{\rm neg}$ & $\sigma$ (pb) & fraction (\%) \\
\hline
$aa2j$  &  $3.98 \pm 0.01$ & 0.39 &  $0.87 \pm 0.02$ &  2 \\
$aa1j$  & $13.99 \pm 0.04$ & 0.21 &  $8.24 \pm 0.06$ & 17 \\
$aa0j$  & $10.33 \pm 0.03$ &  0   & $10.33 \pm 0.03$ & 21 \\
$a2j$   & $10.28 \pm 0.05$ & 0.32 &  $3.76 \pm 0.06$ &  8 \\
$a1j$   & $12.48 \pm 0.06$ &  0   & $12.48 \pm 0.06$ & 26 \\
$qcd2j$ &  $4.94 \pm 0.04$ &  0   &  $4.94 \pm 0.04$ & 10 \\
$gg2aa$ &  $7.74 \pm 0.02$ &  0   &  $7.74 \pm 0.02$ & 16 \\
\hline
total   &  &  & $48.36 \pm 0.11$ &  \\
\hline
\end{tabular}
\end{center}
\end{table}

\begin{figure}[tp]
\begin{center}
\includegraphics[scale=0.8]{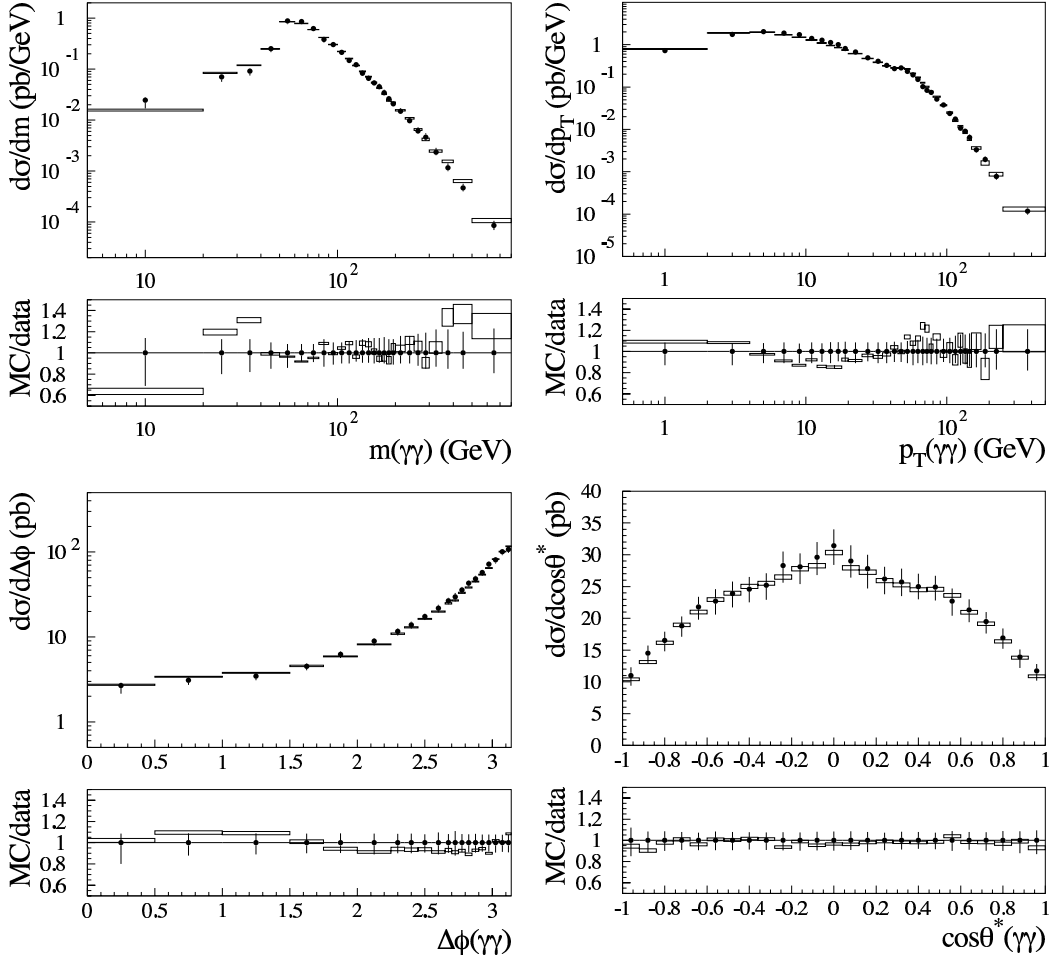}
\caption{\label{fig:comp-atlas}
Comparison with the ATLAS measurement~\cite{Aad:2012tba}.
The results for four quantities of the two-photon system, 
the invariant mass ($m$), transverse momentum ($p_{T}$), azimuthal-angle 
separation ($\Delta\phi$), and production angle in the Collins-Soper frame 
($\theta^{*}$) are compared, and both the direct comparison and ratio are displayed.
The measurement results are plotted using filled circles with attached error bars.
The error bars represent the total error of the measurement presented in the report.
The simulation results are shown with boxes.
The vertical size of the boxes represents the statistical error of the simulation.
The simulation results are multiplied by an overall factor of 0.910.
}
\end{center}
\end{figure}

The event simulation was carried out for the tree-level matched diphoton processes 
in Section~\ref{sec:strategy} with the standard setting of $\mu = \mu_{0}$ 
and the $gg \rightarrow \gamma\gamma$ process described in Section~\ref{sec:gg2gamgam}.
The simulation results were obtained by adding all these simulations. 
The resultant contribution of each process is summarized in Table~\ref{table:atlas}.
We can see that no process dominates the result.
Although the contribution of $aa2j$ seems to be relatively small, 
its contribution to the differential cross section is substantial 
because the negative-weight fraction is large.
Here, $\sigma_{\rm abs}$ is the integration of the absolute value of 
the differential cross section, {\it i.e.}, 
\begin{equation}\label{eq:xsec_abs}
   \sigma_{\rm abs} = \int \left| { d\sigma \over d\Phi} \right| d\Phi,
\end{equation}
where $\Phi$ represents the phase space.
Events are generated in proportion to $\left| d\sigma / d\Phi \right|$ 
and an event weight of $+1$ or $-1$ is assigned according to the sign 
of $d\sigma / d\Phi$.
The negative-weight fraction $R_{\rm neg}$ is the fraction of events that have 
the event weight of $-1$.
Thus, the true cross section can be obtained as 
\begin{equation}\label{eq:xsec_true}
   \sigma = \left( 1 - 2 R_{\rm neg} \right) \sigma_{\rm abs} . 
\end{equation}
The same calculation is performed in each measurement bin in order to obtain 
the kinematical distributions.

The simulated kinematical distributions are compared 
with the measurement results in Fig.~\ref{fig:comp-atlas}.
Because LO simulations are not capable of predicting absolute values, 
the total yield of the simulation is normalized to the total cross section 
of $44.0^{+3.2}_{-4.2}$ pb in the ATLAS measurement; 
namely, the simulation results are multiplied by a factor of 0.910.
The results in Fig.~\ref{fig:comp-atlas} show that 
the simulation is in good agreement with the measurement within the measurement errors.
Contrary to the SHERPA simulation examined in the ATLAS report~\cite{Aad:2012tba}, 
we observe no systematic discrepancy significantly larger than the measurement errors.
However, although they are contained within the measurement errors, 
some systematic tendencies can be observed in the $p_{T}$ and $\Delta\phi$ distributions. 

We pursued possible improvements by changing tunable parameters in the simulation.
Our simulation includes only a few parameters that we can optimize.
The overall normalization has already been optimized.
Another tunable parameter is the energy scale $\mu$.
Although the simulation results are very stable against the variation of $\mu$, 
small dependencies remain as a result of multiple radiation in PS~\cite{Odaka:2009qf}. 
As a test, we repeated the simulation of the tree-level matched diphoton 
processes with the settings of $\mu = 0.5 \mu_{0}$ and $1.5 \mu_{0}$ 
as in the matching test.
The simulation of $gg2aa$ was unchanged.
The total cross section increased by 37\% for the setting of $\mu = 0.5 \mu_{0}$ 
and decreased by 11\% for $\mu = 1.5 \mu_{0}$.
However, after renormalizing the total yield, 
these changes resulted in alterations of only a few percent 
in most of the measurement bins, 
although slightly larger alterations were observed in some edge bins of the $m(\gamma\gamma)$ 
and $p_{T}(\gamma\gamma)$ distributions where measurement errors are large.
Thus, they do not significantly change the observed tendencies.

The PYTHIA simulation includes many tunable parameters.
However, most of them are not relevant to the kinematics of high-energy objects.
As far as we know, the simulation in PYTHIA that most significantly affects 
the kinematics is the primordial-$k_{T}$ simulation~\cite{Odaka:2013fb}, 
which simulates the motion of partons inside hadrons 
that cannot be reproduced by PS simulations. 
The default value of the average $k_{T}$ is set to 2.0 GeV.
We examined the effect of this simulation by changing the average value 
to 0 and 4.0 GeV.
The former was tested by deactivating the primordial-$k_{T}$ simulation.
We applied these PYTHIA simulations to all processes including $gg2aa$.
As a result, we observed a 20\%-level alteration in the smallest $p_{T}$ bin 
and nearly a 10\%-level alteration in the largest $\Delta\phi$ bin. 
However, we did not observe any changes significantly larger than the statistical 
errors of the simulation in the other bins.
Substantially, the simulation results are very stable against the variation 
of model parameters in the simulation relative to the measurement errors.

\subsection{CMS measurement}\label{subsec:cms}

\begin{figure}[tp]
\begin{center}
\includegraphics[scale=0.8]{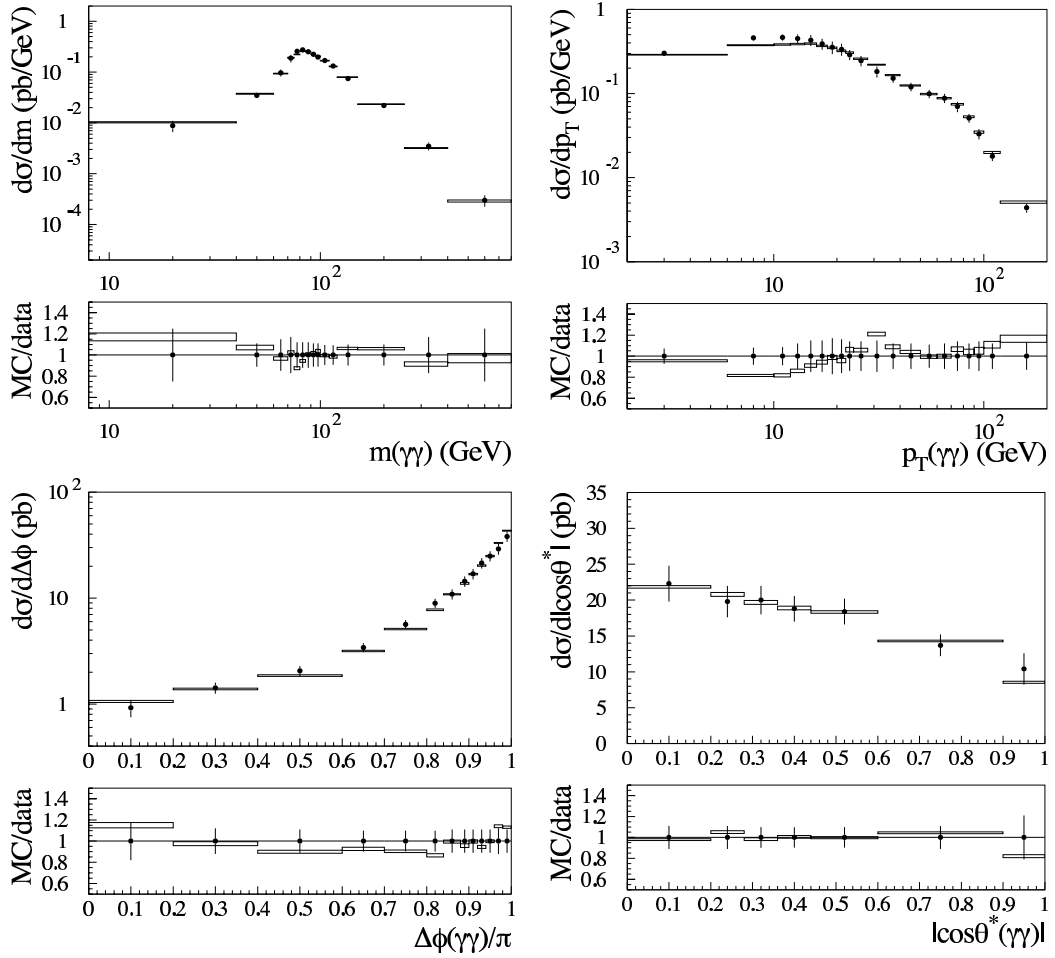}
\caption{\label{fig:comp-cms}
Comparison with the CMS measurement~\cite{Chatrchyan:2014fsa}.
The notations are the same as in Fig.~\ref{fig:comp-atlas}, 
except that the $\theta^{*}$ distribution is presented as a function of 
$\left | \cos\theta^{*} \right |$. 
The simulation results are multiplied by an overall factor of 0.883.
}
\end{center}
\end{figure}

The CMS experiment also performed a measurement regarding diphoton kinematical 
distributions~\cite{Chatrchyan:2014fsa}. 
The measurement was based on 5.0 fb$^{-1}$ data of $pp$ collisions 
at $\sqrt{s}$ = 7 TeV provided by the LHC.
Although the same quantities as those in the ATLAS measurement were measured, 
the signal definition was slightly different.
The angular coverage was defined as 
\begin{equation}\label{eq:phang_cms}
  |\eta({\gamma})| < 1.44 {\rm ~~or~~} 1.57 < |\eta({\gamma})| < 2.5 .  
\end{equation}
The two photons in this angular range were required to satisfy the following conditions, 
\begin{equation}\label{eq:phmom_cms}
  p_{T}(\gamma_{1}) > 40 {\rm ~GeV}/c , {\rm ~~} p_{T}(\gamma_{2}) > 25 {\rm ~GeV}/c , 
  {\rm ~~and~~} \Delta R(\gamma\gamma) > 0.45 .
\end{equation}
The threshold values are larger than in the ATLAS measurement.
Although the photon-isolation condition is not explicitly defined in their report, 
a condition of 
\begin{equation}\label{eq:phiso_cms}
  E_{T}^{\rm cone}(\Delta R < 0.4) < 5 {\rm ~GeV}  
\end{equation}
was required in their simulation studies.

We simulated the measurement following the above conditions, 
again using the MRST2007LO* PDF. 
The UE simulation in PYTHIA was deactivated, and muons and neutrinos were excluded 
from the calculation of $E_{T}^{\rm cone}$, 
as in the simulation for the ATLAS measurement. 
The obtained simulation results are compared with the measurement 
in Fig.~\ref{fig:comp-cms}.
The total cross section was measured to be $17.2 \pm 2.0$ pb by CMS, 
while the cross section that we obtained from the simulation is 19.48 pb.
Hence, the simulation results are multiplied by a factor of 0.883.
This normalization factor is consistent with that for the ATLAS measurement. 
The difference is well within the measurement errors.

In Fig.~\ref{fig:comp-cms}, 
we again observe overall good agreement between the simulation and measurement.
However, the agreement is marginal in the $p_{T}(\gamma\gamma)$ distribution.
A significant deficit of the simulation can be observed around 
$p_{T}(\gamma\gamma)$ = 10 GeV, and a wavy structure is observed in the ratio.
This structure is similar to the tendency observed in the comparison 
with the ATLAS measurement. 
However, the details of the structure, such as the minimum/maximum positions, 
are different.
The different $p_{T}$ thresholds may have resulted in this difference.

A similar deficit of the simulation was also observed at 
$p_{T}(\gamma\gamma) \lesssim 20$ GeV in the SHERPA simulation examined 
by CMS~\cite{Chatrchyan:2014fsa}, 
where the deficit continuously remains down to $p_{T}(\gamma\gamma) = 0$.
Note that the simulation/measurement ratio is inverted and the SHERPA simulation 
is not normalized to the measurement in the CMS study.
The SHERPA simulation in the CMS paper shows better agreement with the measurement 
in the $\Delta\phi(\gamma\gamma)$ distribution than ours in Fig.~\ref{fig:comp-cms}.
The SHERPA simulation examined by CMS includes 3-jet contributions.
This must be the reason for the better performance 
in the $\Delta\phi(\gamma\gamma)$ distribution, 
whereas it does not result in a better performance in $p_{T}(\gamma\gamma)$.

The CMS measurement applies a very asymmetric $p_{T}$ cut to the photons.
The subleading photon cannot have a $p_{T}$ value smaller than 
the threshold for the leading photon 
if there is no additional transverse activity.
Additional QCD activities allow the subleading photon to have $p_{T}$ values 
in the range between the two thresholds, $p_{T{\rm , min}}^{(1)}$ 
and $p_{T{\rm , min}}^{(2)}$.
As a result, the $p_{T}(\gamma\gamma)$ distribution at 
$p_{T}(\gamma\gamma) < p_{T{\rm , min}}^{(1)} - p_{T{\rm , min}}^{(2)}$ 
is reduced.
Thus, the observed deficit in this range implies that the QCD activity 
in the simulation is too strong 
and the activity in SHERPA is stronger than ours.
This discussion seems to be inconsistent with the observation 
in the $\Delta\phi(\gamma\gamma)$ distribution, 
where SHERPA provides a better simulation. 
However, the relevant $p_{T}$ range of the QCD activity may be different 
for the two quantities.
The observations can be consistently understood if relatively soft activities 
that are simulated by PS are effective for small $p_{T}(\gamma\gamma)$ values, 
while the $\Delta\phi(\gamma\gamma)$ distribution is sensitive to 
harder activities that are simulated according to multi-jet MEs.

\section{Discussion}\label{sec:discuss}

\subsection{Acoplanar diphoton}\label{subsec:acoplanar}

\begin{figure}[tp]
\begin{center}
\includegraphics[scale=0.6]{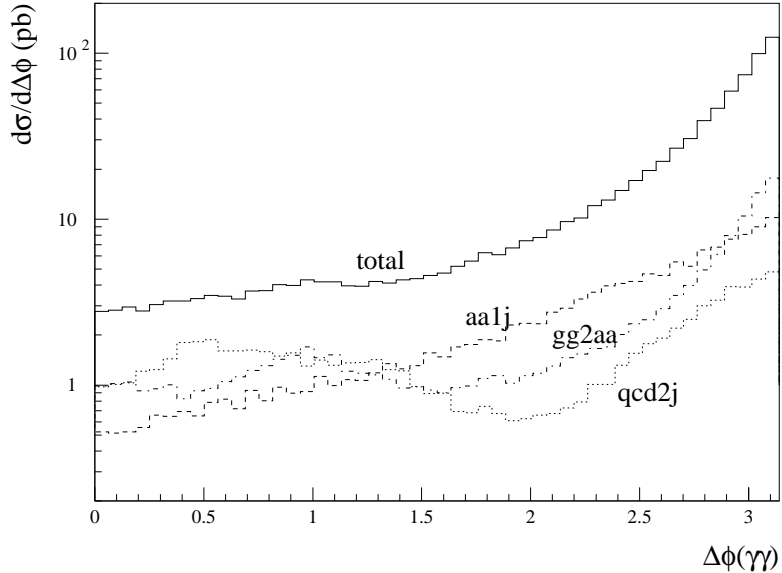}
\caption{\label{fig:dphi_proc}
$\Delta\phi(\gamma\gamma)$ distribution of the simulation for the ATLAS measurement.
Contributions from $aa1j$, $qcd2j$, and $gg2aa$, which dominate the distribution 
at small $\Delta\phi$ are separately displayed together with the sum of all processes.
The results for $qcd2j$ and $gg2aa$ are identical to those in Figs.~\ref{fig:dphi-qcd2j} 
and \ref{fig:gg2aa}, respectively.
}
\end{center}
\end{figure}

In the Tevatron studies, 
the most significant discrepancy between the measurement and NLO predictions 
was observed in the $\Delta\phi(\gamma\gamma)$ distribution at small values.
The $\Delta\phi(\gamma\gamma)$ distribution of our simulation for the ATLAS measurement 
is shown in Fig.~\ref{fig:dphi_proc}.
In the figure, the contributions from $aa1j$, $qcd2j$, and $gg2aa$ processes, 
which dominate the distribution in acoplanar (small $\Delta\phi$) regions, 
are separately presented together with the total sum.
We can see that these three processes have nearly equal importance in acoplanar regions.

Among these three processes, the $aa1j$ contribution decreases 
as $\Delta\phi(\gamma\gamma)$ decreases 
because there is no enhancement of collinear $\gamma\gamma$ production.
This contribution must have been properly included in the NLO predictions 
because $aa1j$ is a process at NLO.
In contrast, as discussed in Section~\ref{subsec:photorad}, 
$qcd2j$ shows an enhancement of collinear production due to the double radiation.
Formally, this process first appears in the NNLO approximation. 
In addition, $gg2aa$ events with a hard radiation should formally be categorized 
as an NNNLO correction.
Therefore, the absence or an inappropriate evaluation of the $qcd2j$ and/or $gg2aa$ 
contributions must have been the reason for the deficit in NLO predictions.

Additional hard radiation effects are necessary to produce acoplanar $gg2aa$ events.
Such effects are simulated by increasing the PS energy scale in our simulation, 
as described in Section~\ref{sec:gg2gamgam}.
Namely, the hard radiation effects are added according to the LL approximation.
Usually, the LL approximation becomes less accurate as the radiation $Q^{2}$ 
becomes comparable with or higher than the typical energy scale of the hard interaction.
An enhancement around $\Delta\phi = 1$ that we can observe in the $gg2aa$ result 
in Figs.~\ref{fig:gg2aa} and \ref{fig:dphi_proc} may have been caused by this inaccuracy.
It would also be necessary to apply the radiation matching method to $gg2aa$ 
if we need more accurate predictions in acoplanar regions.

\begin{figure}[tp]
\begin{center}
\includegraphics[scale=0.8]{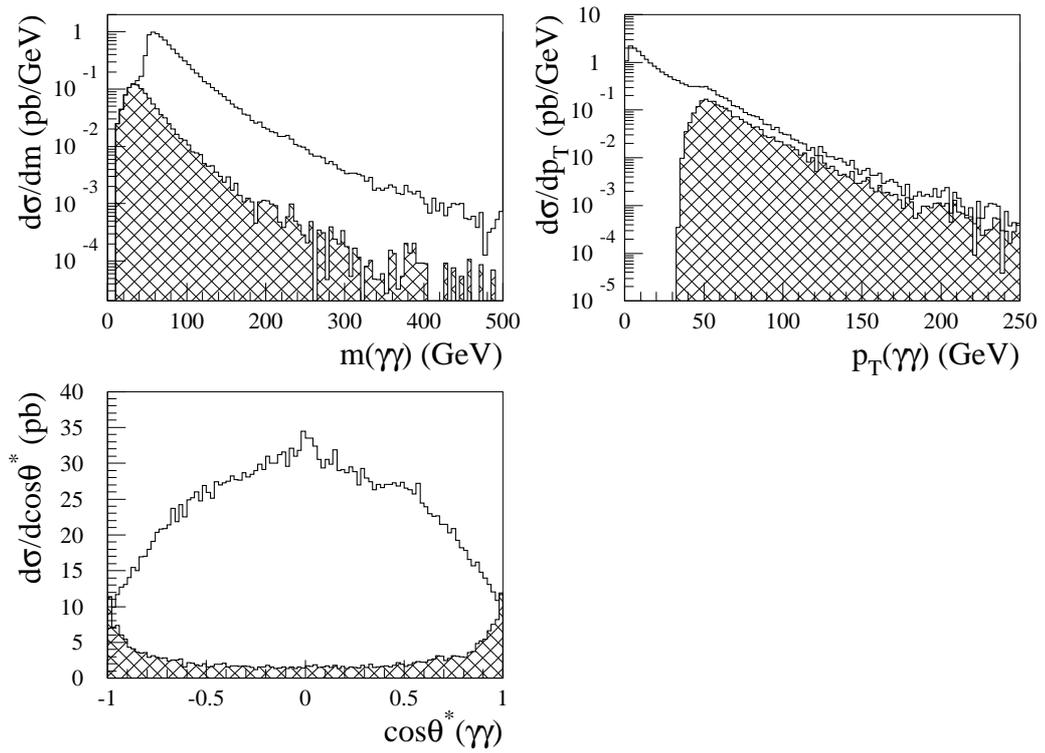}
\caption{\label{fig:acopevt}
Contribution of acoplanar diphoton events in other distributions.
The distributions of events in an acoplanar region, 
$\Delta\phi(\gamma\gamma) < 1.5$, in the simulation for the ATLAS measurement 
are displayed with hatched areas.
}
\end{center}
\end{figure}

The hatched areas in Fig.~\ref{fig:acopevt} show the contribution of events 
in an acoplanar region, $\Delta\phi(\gamma\gamma) < 1.5$, to the other distributions. 
We can see that the low-mass distribution and bump structure in the $p_{T}$ distribution 
are totally determined by these acoplanar events.
In addition, the acoplanar events predominantly determine the $\cos\theta^{*}(\gamma\gamma)$ 
distribution close to $\pm 1$, where all theoretical predictions and simulations 
failed to reproduce the measurements~\cite{Aad:2012tba,Chatrchyan:2014fsa}.
Our simulation reproduces the measurements very well in these regions, 
as we can see in Figs.~\ref{fig:comp-atlas} and \ref{fig:comp-cms}.
However, the scattering angle $\theta^{*}$ in the Collins-Soper frame has been introduced 
to probe the properties of underlying hard interactions.
We are not sure whether the $\theta^{*}$ of acoplanar events is helpful for such studies.
It may be better to apply a certain coplanarity-angle cut 
in the study of the $\theta^{*}$ distribution.

\subsection{Photon isolation}\label{subsec:isolation}

\begin{figure}[tp]
\begin{center}
\includegraphics[scale=0.6]{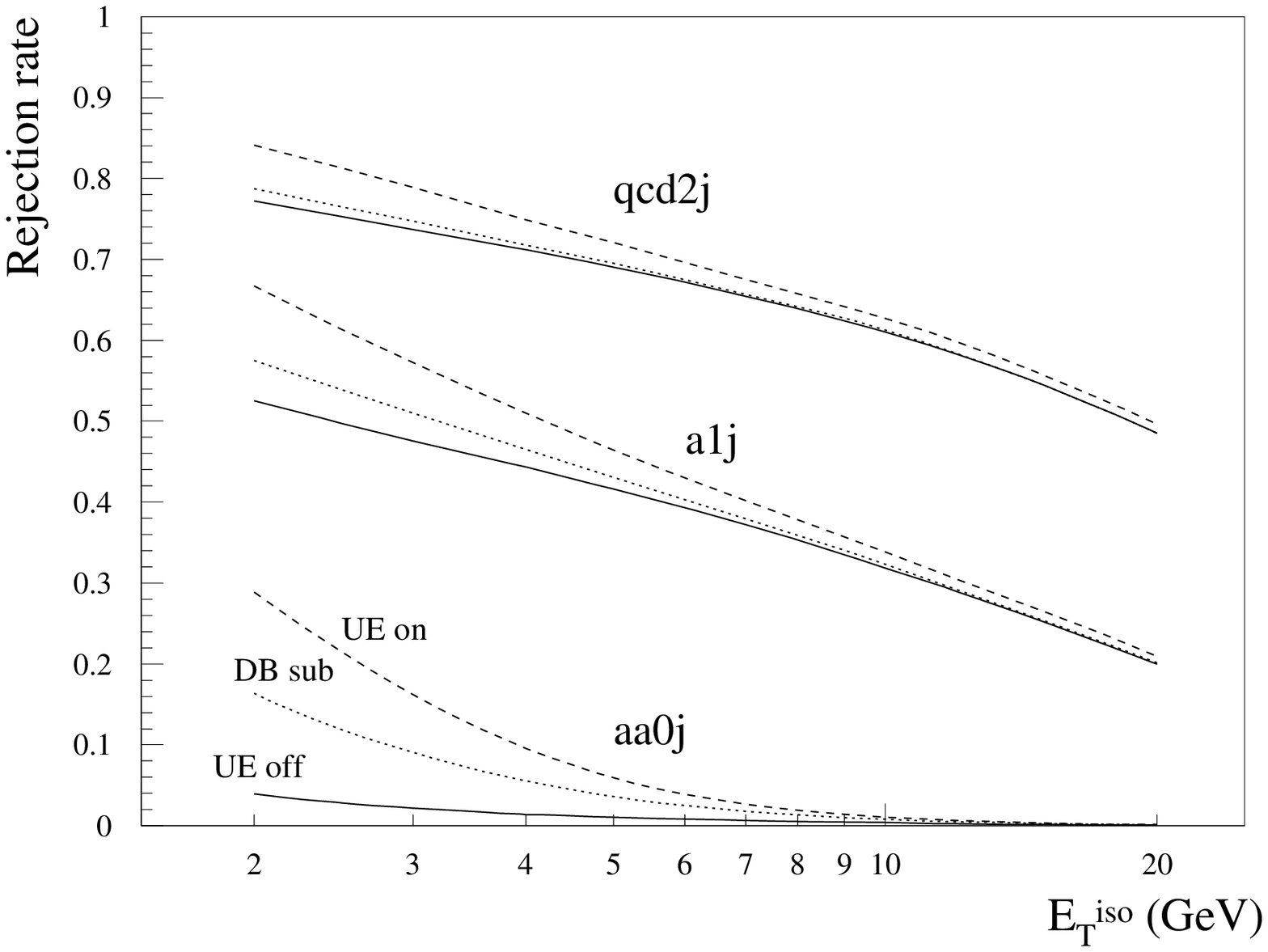}
\caption{\label{fig:isorej}
Rejection rate of the photon isolation cut in the simulation for 
the ATLAS measurement.
The results for three characteristic processes are presented as a function 
of cut value $E_{T}^{\rm iso}$.
The rates obtained in the simulations without and with the underlying-event (UE) 
simulation are illustrated with solid and dashed lines, respectively.
The results obtained with the diffuse-background (DB) subtraction, 
which is described in the text, are represented by dotted lines.
}
\end{center}
\end{figure}

We deactivated the UE simulation in the comparison with the LHC measurements 
because ATLAS and CMS state in their reports that the UE contributions are subtracted 
from the cone $E_{T}$ for the photon isolation and because the UE effects to the 
examined kinematical distributions are negligible.
However, it is not trivial that the quantities simulated by the UE simulation in PYTHIA 
are identical to the quantities that the experiments subtracted.
As a test, we repeated the simulation for the ATLAS measurement by activating the 
default UE simulation in PYTHIA 6.425.

Figure~\ref{fig:isorej} shows the rejection rate of the isolation cut, defined as
\begin{equation}\label{eq:phiso_gen}
  E_{T}^{\rm cone}(\Delta R < 0.4) < E_{T}^{\rm iso} . 
\end{equation}
The results are presented for three typical processes, and those without and with 
the UE simulation are compared as a function of $E_{T}^{\rm iso}$.
The rejection rate is very small in the $aa0j$ events because the produced photons 
do not have any correlation with additional hadronic activities.
In contrast, in the $qcd2j$ events, photons are produced as PS radiations 
from quarks.
Hence, they are likely to be surrounded by hadronic activities induced by the quarks. 
As a result, the rejection rate becomes high.
The $a1j$ results are observed in approximately the middle of them because one of the 
two photons originates from the PS simulation.

We observe in these results that the UE effect is process dependent.
Because the UE particles alter the $E_{T}^{\rm cone}$ value, 
the effect is larger if the $E_{T}^{\rm cone}$ spectrum is steeper.
The effect would be negligible if the spectrum was flat.
In order to examine such a UE effect, 
we applied the UE simulation to all processes and applied the isolation cut 
in Eq.~(\ref{eq:phiso_atlas}) to the simulated events.
In spite of a clear process dependence in Fig.~\ref{fig:isorej}, 
the simulated normalized kinematical distributions did not show any significant 
difference to those in Fig.~\ref{fig:comp-atlas}.
The differences were well contained within the statistical error of the simulations, 
although the total cross section decreased by 10\% with respect to the value 
obtained from the simulation in Section~\ref{subsec:atlas}.

The above is a discussion of extreme cases.
The ATLAS and CMS experiments describe that they subtract the UE contribution, 
together with the pile-up effects, from $E_{T}^{\rm cone}$ event-by-event 
using the jet-area/median method~\cite{Cacciari:2007fd}
implemented in the FastJet package~\cite{Cacciari:2011ma}.
Although the experiments must have applied the subtraction to the detector-level data, 
we tried to apply it to particle-level simulation data 
in order to examine the effect of this method.
As in the jet clustering in Section~\ref{sec:matching}, 
we used all stable particles in the angular range of $|\eta| < 5$ 
in the simulated events to which the UE simulation was applied, 
in order to evaluate the average diffuse-background (DB) $p_{T}$ density ($\rho$) 
on the $\eta$-$\phi$ plane.
The recommended Cambridge/Aachen algorithm with $R$ = 0.6 was used for the jet clustering. 
However, the evaluation frequently failed and returned $\rho$ = 0.
This was because the DB particles were too sparse with the UE simulation only.
Indeed, although the quantity $\zeta = \sigma/(\sqrt{\langle A \rangle}\rho)$ is required to be 
sufficiently smaller than unity in this method~\cite{Cacciari:2007fd}, 
its value was frequently close to unity even when a non-zero $\rho$ was returned.
Here, $\sigma$ and $\langle A \rangle$ denote the internal fluctuation parameter and average 
jet-area, respectively, which are returned by the FastJet program.

We found that this problem can be solved by adding fake particles that imitate 
the pile-up particles.
Here, we call them pseudo-pile-up (PPU) particles.
We added the PPU particles uniformly and randomly on the $\eta$-$\phi$ plane 
within the range of $|\eta| < 5$.
The $p_{T}$ values were determined according to a Gaussian distribution with 
an average of 1.0 GeV/$c$ and root-mean-square of 0.2 GeV/$c$.
A Gaussian distribution was chosen because the average $p_{T}$ density 
($\rho_{\rm PPU}$) is trivial and it is easy to tune the fluctuation.
We also tried to use an exponential distribution that simulates the low-$p_{T}$ 
spectrum of particles in the simulated events.
Although the obtained results were similar, 
we observed a slightly better performance with the Gaussian distribution 
in the $\zeta$-parameter distribution.

We added 500 such PPU particles to each event.
They were assumed to be massless.
From the $\rho$ value returned from the FastJet program, 
we evaluated the UE density as $\rho_{\rm UE} = \rho - \rho_{\rm PPU}$. 
The UE contribution to the isolation-cone $E_{T}$ was estimated to be 
$\rho_{\rm UE}A_{\rm iso}$, 
where $A_{\rm iso} = \pi (\Delta R)^{2}$ with $\Delta R$ = 0.4.
The estimated UE contribution was subtracted from the measured $E_{T}^{\rm cone}$ 
before applying the isolation cut in Eq.~(\ref{eq:phiso_gen}).
The dotted lines in Fig.~\ref{fig:isorej} show the rejection rates for the 
isolation condition thus defined.
We can see that the subtraction cannot totally remove the activities produced 
by the UE simulation of PYTHIA.
In PYTHIA, UE is simulated by low-$p_{T}$ QCD 2-jet production.
The results in Fig.~\ref{fig:isorej} are reasonable because the thus-simulated 
UE particles may partly form non-diffusive jet-like structures.

Although we observed no significant difference in the normalized kinematical 
distributions, 
the simulation with the DB subtraction and the isolation condition 
in Eq.~(\ref{eq:phiso_atlas}) gives a total cross section approximately 
4\% smaller than the value that was obtained in Section~\ref{subsec:atlas}. 
Hence, looking at the UE-on/off and DB-subtraction results, 
we have to consider that the cross section result that the experiment presented 
has an additional 5\%-level uncertainty 
because the corresponding signal definition has an ambiguity of this level.
Although this uncertainty is smaller than the current measurement precision, 
it would be a serious concern in future higher-precision measurements.
This uncertainty substantially arises because the experiments do not explicitly 
define the treatment of the UE effect.
At least when numerical results are presented, the corresponding signal definition 
has to be unambiguously presented by experiments.
The current definitions of the LHC experiments are ambiguous 
because the UE subtraction is not defined using detector-independent physical quantities.
Furthermore, because UE is a part of the $pp$ interaction, 
it may not be reasonable to exclude its contribution from the definition 
of the isolation condition.
We need further discussion to reach a reasonable consensus on this point.

\subsection{Two-jet contribution}\label{subsec:twojets}

\begin{figure}[tp]
\begin{center}
\includegraphics[scale=0.8]{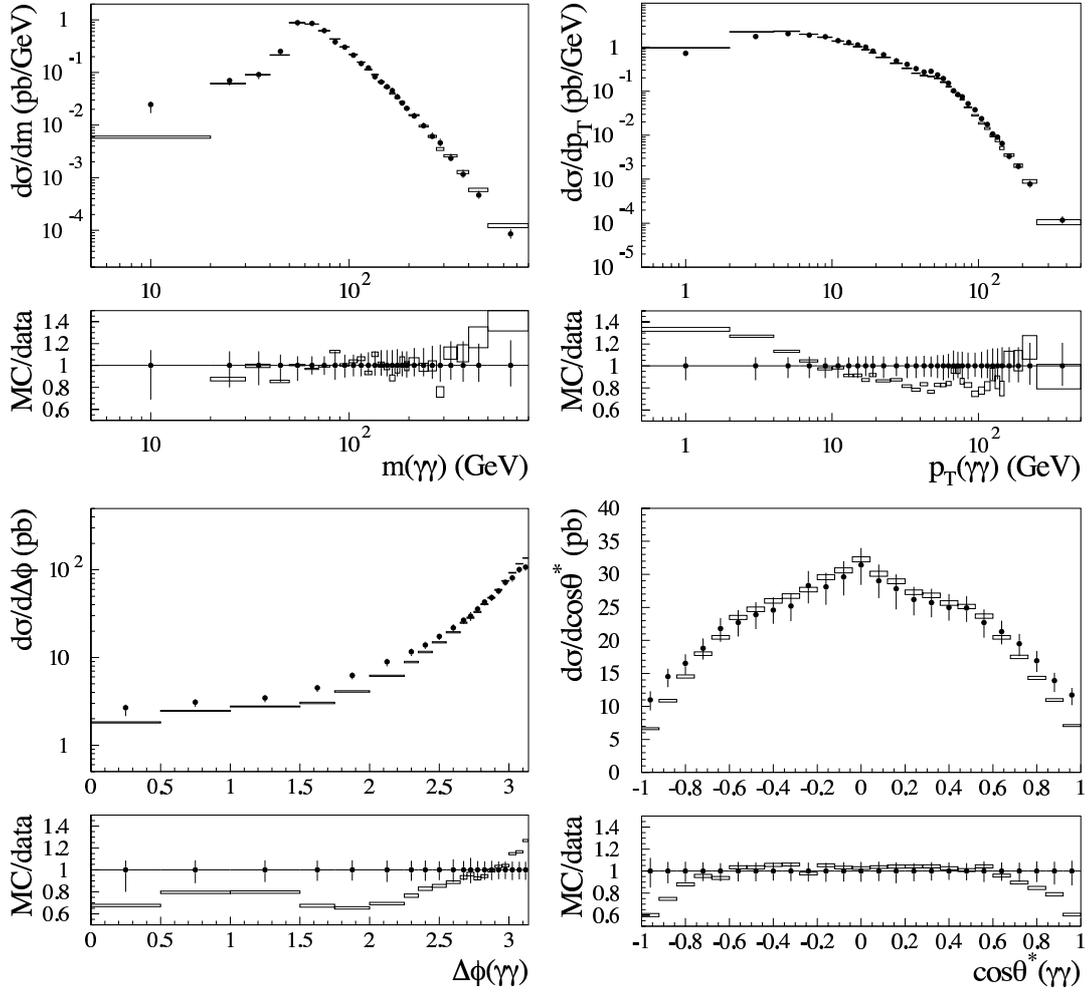}
\caption{\label{fig:upto1j}
Comparison with the ATLAS measurement, as in Fig.~\ref{fig:comp-atlas}, 
except that the simulation results are obtained 
by excluding the 2-jet processes, {\it i.e.}, they are the sum of the $aa1j$, 
$aa0j$, $a1j$, and $gg2aa$ results. 
The soft-gluon correction is not applied to the $aa1j$ and $a1j$ processes.
The simulation results are normalized to the measured total cross section 
by applying a normalization factor of 1.11.
}
\end{center}
\end{figure}

As discussed in the introduction, 
we suspected that the contribution of 2-jet production processes would 
be large in diphoton production 
based on the observation in direct-photon production~\cite{Odaka:2015uqa}. 
From the results in Table~\ref{table:atlas}, 
we can evaluate the 2-jet contribution $aa2j$ + $a2j$ + $qcd2j$ 
to be approximately 20\% of the total yield in our standard simulation 
for the ATLAS measurement.
This value is not small, but it is also not very large 
in the aspect of the QCD correction.
Furthermore, the contribution of new processes, which we suspected to be large, 
is relatively small, as already pointed out in a study of the NNLO 
approximation~\cite{Catani:2011qz}.
The contribution is approximately 3\% from $gg \rightarrow \gamma \gamma q \bar{q}$ 
and 5\% from $qq' \rightarrow \gamma \gamma qq'$ ($q' \neq \bar{q}$) 
in our simulation.
The remainder is dominated by the ordinary gluon-radiation correction to 
$qg \rightarrow \gamma \gamma q$.
Its contribution amounts to approximately 10\%.
The contribution of $q \bar{q}$ initial-state processes is again very small (1.8\%).
Among them, the $q \bar{q} \rightarrow \gamma \gamma gg$ contribution is only 0.3\%.
The remainder comes from other new processes: 
1.0\% from $q \bar{q} \rightarrow \gamma \gamma q \bar{q}$ and
0.5\% from $q \bar{q} \rightarrow \gamma \gamma q' \bar{q}'$ ($q' \neq q$).

Simulation results show that the limited 2-jet contribution is partly due to 
the requirement regarding the photon isolation.
As we can expect from interaction diagrams, 
2-jet processes, particularly the new processes, have larger rejection probabilities 
than the other processes.
If we remove the isolation requirement, the $gg \rightarrow \gamma \gamma q \bar{q}$ 
and $qq' \rightarrow \gamma \gamma qq'$ contributions increase to 6\% and 11\%, 
respectively, and the total 2-jet contribution increases to 34\%.
This is comparable with the 1-jet ($aa1j + a1j$) contribution of 41\%.

Although the impact of 2-jet processes on the total cross section is not very large 
under the actual measurement condition, 
they have a remarkable contribution to kinematical distributions of the two photons.
The simulation results in which the 2-jet processes are excluded are compared with 
the ATLAS measurement results in Fig.~\ref{fig:upto1j}. 
These results should be compared with those in Fig.~\ref{fig:comp-atlas}.
The total yield of the simulation is again normalized to the measured total cross section. 
From these results, we observe that the contribution of the 2-jet processes depends 
on the examined quantity.

The $m(\gamma \gamma)$ distribution of this up-to-1-jet simulation is already 
in good agreement with the measurement, except for the lowest-mass bin.
In contrast, significant discrepancies are observed in the other distributions.
This is reasonable because $m(\gamma \gamma)$ is determined by hard interactions 
to produce the two photons, 
while the $p_{T}(\gamma \gamma)$ and $\Delta\phi(\gamma \gamma)$ distributions are 
sensitive to additional QCD radiations.
The obvious deficit of the simulation in small $\Delta\phi(\gamma \gamma)$ regions 
is caused by the absence of $qcd2j$ events, 
as discussed in Section~\ref{subsec:acoplanar}.
This deficit results in the discrepancy near $\left| \cos \theta^{*} \right| = 1$ 
in the $\cos \theta^{*} (\gamma \gamma)$ distribution.
The up-to-1-jet simulation is already in good agreement with the measurement 
in central regions because $\cos \theta^{*} (\gamma \gamma)$ is predominantly 
determined by hard interactions. 
The remarkable improvement in the $p_{T}(\gamma \gamma)$ distribution at high $p_{T}$, 
$20 \lesssim p_{T}(\gamma \gamma) \lesssim 100$ GeV/$c$, 
indicates a significant contribution of hard radiation effects in the 2-jet processes.
The reason for the remaining deficit around $p_{T}(\gamma \gamma)$ = 10 GeV/$c$ 
must be different from that for the deficit at small $p_{T}(\gamma \gamma)$ 
that we observed in Section~\ref{subsec:cms}, 
because the difference between the two $p_{T}$ thresholds is only 3 GeV/$c$ 
in the ATLAS measurement.

As we discussed above, 
the 2-jet production processes have a large contribution to diphoton production, 
at least when we remove the photon-isolation requirement.
However, contrary to the initial conjecture, 
the large contribution is not predominantly caused by the emergence of new processes.
The largest contribution comes from the gluon-radiation correction to 
$qg \rightarrow \gamma \gamma q$.
A large 2-jet contribution was also observed in direct-photon production 
in our previous study~\cite{Odaka:2015uqa}.
Reexamining the simulation results, we found that the underlying properties are 
almost the same as for diphoton production.
While the 2-jet contribution amounts to 42\% of the total cross section, 
the $gg \rightarrow \gamma q \bar{q}$ and $qq' \rightarrow \gamma qq'$ contributions 
are only 3\% and 7\%, respectively, under the condition applied in Section~4.1 
of the previous paper~\cite{Odaka:2015uqa}.
The $qg \rightarrow \gamma qg$ contribution amounts to approximately 30\%. 
This corresponds to a more than 50\% correction to the dominant lowest-order process 
$qg \rightarrow \gamma q$.

An incomprehensible behavior has also been observed in ordinary perturbative 
calculations for diphoton production~\cite{Catani:2011qz,Campbell:2016yrh}. 
It was found that the energy-scale dependence increases when the approximation 
is improved from NLO to NNLO.
Namely, the naive perturbative nature seems to be violated at NNLO, 
which is the approximation order including associated 2-jet production. 
This strange behavior may be related to the large gluon-radiation contribution
in 2-jet processes.

Although the large gluon-radiation contribution is still a puzzle, 
the large 2-jet contribution can be considered to be a common property of 
photon-production processes in hadron collisions.
Furthermore, this property must also be common to other electroweak-boson production 
processes, such as single $W$ and $Z$ production and diboson 
($W^{+}W^{-}$, $W^{\pm}Z$, and $ZZ$) production, 
because the underlying QCD structure is identical. 
Indeed, we observed a large 2-jet contribution to $W$-boson production 
at high $p_{T}$ in a previous study~\cite{Odaka:2014ura}.
The 2-jet contribution is not significant and thus the NLO approximations provide 
good predictions in low-$p_{T}$ weak-boson production, 
probably because the amplitudes of final-state radiation diagrams, 
such as those corresponding to the diagrams in Fig.~\ref{fig:diagrams}, 
are suppressed by the weak-boson mass.

If the above discussion is also valid for high-mass diboson production, 
we need to take special care regarding the collinear jet production 
in association with weak-boson production, 
especially when the weak bosons are detected in hadronic-decay modes 
as highly-boosted massive objects.
It must be difficult to separate the associated jet from the decay products  
and the weak-boson momenta may be overestimated in such studies.
The probability of collinear jet production may be large 
because approximately 35\% of the events are rejected by the isolation requirement 
in diphoton production under the ATLAS measurement condition.
It would hence be necessary to carry out simulation studies 
with appropriate event generators including associated multi-jet production.

\section{Conclusion}\label{sec:concl}

We have developed an event generator for diphoton ($\gamma\gamma$) production 
in hadron collisions that includes associated jet production up to two jets 
within the framework of the GR@PPA event generator.
Processes having different jet and photon multiplicities are combined 
using a subtraction method based on the LLL subtraction.
The subtracted divergent components in radiation-including processes are consistently 
restored by combining lower multiplicity processes to which PS simulations are applied.

The PS simulation involves QED photon radiations from final-state quarks 
to restore subtracted final-state QED collinear components. 
Photon radiations in very small $Q^{2}$ regions that the PS simulation cannot cover 
are simulated by employing a fragmentation function (FF).
This PS/FF simulation has the ability to enforce energetic photon radiations 
for efficient event generation. 
In principle, the radiation of any number of photons can be enforced in this simulation.
The generated parton-level events can be fed to the PYTHIA event generator 
to obtain particle-level event information.
We can perform realistic photon isolation and hadron-jet reconstruction simulations 
using the obtained events.

The simulated events, in which the loop-mediated $gg \rightarrow \gamma\gamma$ 
process is also involved, 
reasonably reproduce the diphoton production kinematics measured at the LHC. 
The remaining small discrepancies in the azimuthal opening angle 
of the diphoton system indicate the necessity of further higher-order processes.
A question still remains regarding the small discrepancies 
in the $p_{T}(\gamma \gamma)$ distribution.

We found that the contribution of 2-jet processes is significant in diphoton production, 
even in the production kinematics of the two photons.
Contrary to the initial conjecture, 
the contribution of new processes $gg \rightarrow \gamma\gamma q\bar{q}$ and 
$qq' \rightarrow \gamma\gamma qq'$ that first emerge in 2-jet production is not very large.
The largest contribution comes from gluon-radiation corrections to 
$qg \rightarrow \gamma\gamma q$.
Requirements on the photon isolation reduce the 2-jet contribution, 
especially those from the new processes.

The significant 2-jet contribution can be considered as a common property 
of photon-production processes in hadron collisions. 
It may also be common to high-$p_{T}$ weak-boson production.
We need to be careful about the contamination of collinearly produced hadron jets 
when weak bosons are identified as boosted massive jets.

The diphoton production associated with two jets is not yet well understood.
We cannot reasonably explain why the gluon-radiation corrections are large, 
and the naive perturbative nature does not seem to hold in ordinary perturbative 
calculations at NNLO.
These two observations may be related to each other.
We still need further studies in order to improve our understanding.

Improvements in the measurements are also necessary for better understanding.
The treatment of the underlying events in the photon-isolation condition may become 
an obstacle to further improvement.
The contribution from the underlying events is subtracted in the LHC measurements, 
but the subtraction is not defined in a detector-independent form.
The ambiguity in the definition may lead to a cross section uncertainty 
at the level of $\pm 5$\%.
We need to obtain a reasonable consensus concerning the unambiguous signal 
definition.

\section*{Acknowledgments}

This work has been carried out as an activity of the NLO Working Group, 
a collaboration between the Japanese ATLAS group and the numerical analysis 
group (Minami-Tateya group) at KEK.
The authors wish to acknowledge useful discussions with the members.
The programs for the $gg \rightarrow \gamma\gamma$ process was coded by 
Y.~Komori of the University of Tokyo.


\end{document}